\documentstyle[aps,prl,preprint]{revtex}
\begin{document}

\title{Intrinsic State for an extended version of the interacting
  boson model (IBM-4)}
\author{
  J.E.~Garc\'{\i}a--Ramos$^1$,
  J.M.~Arias$^1$,
  J.~Dukelsky$^2$,
  and P.~Van Isacker$^3$ }
\address{$^1$Departamento de F\'{\i}sica At\'omica, Molecular y Nuclear,
  Universidad de Sevilla, Apartado 1065, 41080 Sevilla, Spain}
\address{$^2$Instituto de Estructura de la Materia,
  Serrano 123, 28006 Madrid, Spain}
\address{$^3$Grand Acc\'el\'erateur National d'Ions Lourds,
  B.P.~5027, F-14076 Caen Cedex 5, France}
\date{\today}
\maketitle

\begin{abstract}
  An intrinsic-state formalism for IBM-4 is presented. 
  A basis of deformed bosons is introduced
  which allows the construction
  of a general trial wave function
  which has Wigner's spin--isospin SU(4) symmetry
  as a particular limit.
  Intrinsic-state calculations are compared with exact ones
  showing good agreement.
\end{abstract}

\newpage

The Interacting Boson Model (IBM)
was originally proposed to describe
collective low-lying states in even-even nuclei.
The model building blocks
are monopolar ($s$) and quadrupolar ($d$) bosons.
In the original formulation of the model (IBM-1)
no distinction was made
between neutrons and protons~\cite{Iach87}.
Later, connections with the nuclear shell model
were investigated~\cite{Otsu78b,Iach87b}
and a new version was proposed
in terms of neutron ($s_\nu,d_\nu$)
and proton ($s_\pi,d_\pi$) bosons,
known as IBM-2~\cite{Iach87}.
The model has been widely applied
to medium-mass and heavy nuclei,
where neutrons and protons
are filling different major shells.
In lighter nuclei with $N\approx Z$, however,
neutrons and protons are in the same shell
and a boson made of one neutron and one proton
(known as a $\delta$ boson) should be included.
This version of the boson model,
called IBM-3~\cite{Elli80},
is the simplest isospin invariant formulation of IBM.
The three types of bosons ($\nu$, $\pi$, and $\delta$)
form an isospin $T=1$ triplet
and correspond, microscopically,
to spatially symmetric nucleon pairs with $S=0$.
In particular, the $\delta$ boson corresponds
to a spatially symmetric $S=0$ neutron-proton pair.
A further extension of the IBM
introduces the neutron-proton boson with $T=0$
or $\sigma$ boson,
corresponding to a spatially symmetric nucleon pair
with $S=1$.
This version is known as IBM-4~\cite{Elli81}
and gives a proper description
of even-even as well as odd-odd $N\approx Z$ nuclei. 

The IBM-3 and IBM-4 are appropriate models
for $N\approx Z$ nuclei approaching the proton drip line.
Such nuclei are studied intensively at the moment
in particular with radioactive nuclear beams.
Also, the IBM-4 is a reasonably simple,
yet detailed model
to study the competition between $T=0$ and $T=1$ pairing,
one of the hot topics in current-day nuclear structure physics.

All versions of IBM are algebraic in nature
and do not have a direct geometrical interpretation.
Such interpretation can be achieved, however,
by introducing an intrinsic state
which provides a connection to geometric models
such as that of Bohr and Mottelson~\cite{Bohr75}.
Intrinsic states have been proposed
for IBM-1~\cite{Gino80,Gino80b,Diep80,Duke84}, 
for IBM-2~\cite{Levi90b,Gino92,Pitt85},
and for IBM-3~\cite{Gino94,Garc98}. 
Their primary use
is to provide a geometric visualization
of the model.
In addition, a considerable reduction is achieved
in the complexity of calculations,
which leaves room for the inclusion
of extra degrees of freedom.

The purpose of this letter is
to propose an intrinsic state for IBM-4.
In the limit of strong isovector pairing
it reduces to the intrinsic state for IBM-3;
in general, it can be used for studying the competition
between $T=0$ and $T=1$ pairing
in $N\approx Z$ nuclei.
First, the mean-field formalism for IBM-4 is presented.
This formalism is subsequently checked against
the results of an exact calculation.

The ensemble of bosons in the IBM-4
consists of isovector $T=1$ and isoscalar $T=0$ bosons
which have intrinsic spin $S=0$ and $S=1$, respectively,
to ensure spatial symmetry.
The allowed spin-isospin combinations
are thus $(T,S)=(1,0)$ and $(T,S)=(0,1)$.
These, together with the orbital angular momenta $\ell=0,2$,
give rise to 36 different bosons.
The corresponding boson creation and annihilation operators
are $\gamma^\dag_{\ell m,T\tau,S\sigma}$
and $\gamma_{\ell m,T\tau,S\sigma}$
where $\ell$ is the orbital angular momentum,
$m$ is its projection,
$T$ is the isospin,
$\tau$ is its projection,
$S$ is the spin,
and $\sigma$ is its projection. 
The operators 
$\tilde{\gamma}_{\ell m,T\tau,S\sigma}=
(-1)^{\ell+T+S-m-\tau-\sigma}
\gamma_{\ell -m,T-\tau,S-\sigma}$
are introduced for having appropriate tensor
transformation properties.

The construction of an intrinsic state requires two ingredients.
First, it needs a basis of deformed bosons
and secondly, it requires a trial wave function.
The deformed bosons are defined 
in terms of the spherical ones
through a unitary Hartree-Bose transformation,
\begin{equation}
\label{Har1-ibm4}
\Omega _{p, T \tau, S \sigma }^\dagger=
\sum_{\ell m}\lambda_{\ell m}^{p T \tau S \sigma}
\gamma_{\ell m,T \tau,S \sigma}^\dagger,
\qquad
\gamma_{\ell m,T \tau,S\sigma }^\dagger=
\sum_p\lambda_{\ell m}^{p T \tau S \sigma *}
\Omega_{p,T \tau,S\sigma}^\dagger,
\end{equation}
and their hermitian conjugates.
The deformation parameters $\lambda$ in these equations
verify the following orthonormalization relations:
\begin{equation}
\label{orto-ibm4}
\sum_{\ell m}
\lambda_{\ell m}^{p'T\tau S \sigma *}
\lambda_{\ell m}^{p T \tau S \sigma}
=\delta_{pp'},
\qquad
\sum_{p}
\lambda_{\ell m}^{p T \tau S \sigma *}
\lambda_{\ell'm'}^{p T \tau S \sigma}
=\delta_{\ell\ell'}\delta_{mm'}.
\end{equation}
For convenience, a global label $\xi$ is used
for spin, isospin, and their projections,
$\xi\equiv(T\tau S\sigma)$.
This new index plays the same role in IBM-4
as the isospin projection $\tau$ does
in the IBM-3 intrinsic-state formalism~\cite{Garc98}.
The index $p$ labels the different deformed bosons.
The fundamental deformed boson has $p=0$
while excited ones have
$p=1,2,\ldots,\sum_\ell (2\ell+1)-1$.
If only $s$ and $d$ bosons are included,
the maximum value of $p$ is 5.
Only the ground-state properties are considered here;
so the superscript $p$ is always zero
and can be omitted henceforth.

The definition of the ground-state trial wave function
follows Ref.~\cite{Garc98};
it is different depending on whether the system
is even-even and odd-odd.
For even-even nuclei with proton excess
(the case of neutrons excess is obtained
by interchanging $N_p$ and $N_n$)
the proposed trial wave function
for the ground state has the form
\begin{equation}
\label{Gswf-ibm4-pp}
\left|\phi(\delta,\alpha_T,\alpha_S)\right\rangle_{ee}
={\Delta^\dagger}^{N_n}(\delta,\alpha_T,\alpha_S)
\Omega_{T=1\tau=1}^{\dagger N_p-N_n}
\left|0\right\rangle,
\end{equation}
where $N_n$ ($N_p$)
is half the number of valence neutrons (protons)
and
\begin{eqnarray}
\label{pair-delta}
\Delta^\dagger(\delta,\alpha_T,\alpha_S)&=&
(\Omega^\dag_{T=1\tau=1}\Omega^\dag_{T=1\tau=-1}+
\alpha_T\Omega^\dag_{T=1\tau=0}\Omega^\dag_{T=1\tau=0})
\nonumber\\
&+&\delta
(\Omega^\dag_{S=1\sigma=1}\Omega^\dag_{S=1\sigma=-1}+
\alpha_S\Omega^\dag_{S=1\sigma=0}\Omega^\dag_{S=1\sigma=0}).
\end{eqnarray}
The description of odd-odd nuclei
is complicated even in the ground state,
since in general its spin-isospin values
are not known {\it a priori}.
In $N=Z$ nuclei, which is the case of most interest,
those values are known,
being either $(T,S)=(1,0)$ or $(T,S)=(0,1)$.
Correspondingly, two trial wave functions are proposed
\begin{equation}
\label{Gswf-ibm4-ii1}
\left|\phi(\delta,\alpha_T,\alpha_S)\right\rangle_{oo-1}
={\Delta^\dagger}^{N_n-{1\over 2}}(\delta,\alpha_T,\alpha_S)
{\Omega_{T=1\tau=1}^\dagger}
\left|0\right\rangle,
\end{equation}
and  
\begin{equation}
\label{Gswf-ibm4-ii2}
\left|\phi(\delta,\alpha_T,\alpha_S)\right\rangle_{oo-2,\sigma}
={\Delta^\dagger}^{N_n-{1\over 2}}(\delta,\alpha_T,\alpha_S)
{\Omega_{S=1\sigma}^\dagger}
\left|0\right\rangle.
\end{equation}
Which of these two states is lower in energy
depends on a delicate balance of the different terms
in the Hamiltonian
which in turn follow from the competition
between $T=0$ and $T=1$ pairing.

In addition to the deformation parameters,
three variational parameters,
$\alpha_T$, $\alpha_S$, and $\delta$, 
appear in the trial wave functions. 
The first two are related to
isospin and spin symmetry breaking
in the trial wave function.
For deformation parameters
independent of $\xi$ and for $\alpha_T=\alpha_S=-1/2$,
the operator $\Delta^\dagger(\delta,\alpha_T,\alpha_S)$
corresponds to a bosonic pair
scalar in spin and isospin.
The parameter $\delta$
measures the relative importance
of isovector and isoscalar bosons
in the ground state.
In the limit of $\delta=0$,
the number of isoscalar bosons
in the ground state is zero
and the IBM-3 trial state is recovered~\cite{Garc98}.
Another interesting limit is $\delta=\pm 1$
which is obtained if the IBM-4 Hamiltonian
has Wigner's SU(4) symmetry~\cite{Elli81}.
In this case $T=0$ and $T=1$ bosons
are treated on equal footing. 

Given a general IBM-4 Hamiltonian, $\hat H$,
the ground-state equilibrium parameters
are obtained by minimizing
the expectation value of the Hamiltonian
in the appropriate trial wave
function~(\ref{Gswf-ibm4-pp},\ref{Gswf-ibm4-ii1},\ref{Gswf-ibm4-ii2}). 
A general expression for this expectation value
can be written as
\begin{eqnarray}
\label{Ener-ibm4}
E(\lambda,\delta,\alpha_T,\alpha_S)&=&
\sum_{\xi_1\xi_2}\epsilon_{\xi_1\xi_2} 
\tilde {f_1}(\delta,\alpha_T,\alpha_S,\xi_1\xi_2)\nonumber\\
&+&\sum_{\xi_1\xi_2\xi_3\xi_4}
V_{\xi_1,\xi_2,\xi_3,\xi_4}^c
\tilde{f_2}(\delta,\alpha_T,\alpha_S,\xi_1\xi_2\xi_3\xi_4),
\end{eqnarray}
where,
\begin{equation}
\label{Htauinf1-ibm4}
\epsilon_{\xi_1,\xi_2}=
\sum_{\ell_1 m_1 \ell_2 m_2 }
\tilde \varepsilon_{\ell_1 m_1 \xi_1 \ell_2 m_2 \xi_2}
\lambda_{\ell_1 m_1}^{\xi_2 *}
\lambda_{\ell_2 m_2}^{\xi_2},
\end{equation}
\begin{equation}
\label{Htauinf2-ibm4}
\begin{array}{c}
V_{\xi_1,\xi_2,\xi_3,\xi_4}^c=
\sum\limits_{\ell_1m_1\ell_2m_2\ell_3m_3\ell_4m_4}
{\displaystyle
{V_{\ell_1m_1\xi_1,\ell_2m_2\xi_2,\ell_3m_3\xi_3,\ell_4m_4\xi_4}}}
~ \lambda_{\ell_1m_1}^{\xi_1 *}
\lambda_{\ell_2m_2}^{\xi_2 *}
\lambda_{\ell_3m_3}^{\xi_3}
\lambda_{\ell_4m_4}^{\xi_4},
\\
\end{array}
\end{equation}
\begin{equation}
\label{f1-ibm4}
\tilde{f_1}(\delta,\alpha_T,\alpha_S,\xi_1\xi_2)=
\delta_{\xi_1\xi_2}
\frac{\left\langle\phi(\delta,\alpha_T,\alpha_S)
|\Omega_{\xi_1}^\dagger\Omega_{\xi_2}|
\phi(\delta,\alpha_T,\alpha_S)\right\rangle}
{\langle\phi(\delta,\alpha_T,\alpha_S)\mid
\phi(\delta,\alpha_T,\alpha_S)\rangle},
\end{equation}
and
\begin{equation}
\label{f2-ibm4}
\tilde{f_2}(\delta,\alpha_T,\alpha_S,\xi_1\xi_2\xi_3\xi_4)=
\frac{\langle\phi(\delta,\alpha_T,\alpha_S)
|\Omega_{\xi_1}^\dagger\Omega_{\xi_2}^\dagger
\Omega_{\xi_3}\Omega_{\xi_4}|
\phi(\delta,\alpha_T,\alpha_S)\rangle}
{\langle\phi(\delta,\alpha_T,\alpha_S)\mid
\phi(\delta,\alpha_T,\alpha_S)\rangle}.
\end{equation}
The isospin matrix elements~(\ref{f1-ibm4},\ref{f2-ibm4})
are calculated straightforwardly
from a binomial expansion of the trial wave function.
Furthermore, the parameters
$\tilde\varepsilon_{\ell_1 m_1 \xi_1 \ell_2 m_2 \xi_2}$
and $V_{\ell_1m_1\xi_1,\ell_2m_2\xi_2,\ell_3m_3\xi_3,\ell_4m_4\xi_4}$
in equations~(\ref{Htauinf1-ibm4},\ref{Htauinf2-ibm4})
are defined as
\begin{equation}
\tilde \varepsilon_{\ell_1 m_1 \xi_1 \ell_2 m_2 \xi_2}\equiv
\langle\ell_1m_1\xi_1|\hat H|\ell_2m_2\xi_2\rangle,
\end{equation}
\begin{eqnarray}
V_{\ell_1m_1\xi_1,\ell_2m_2\xi_2,\ell_3m_3\xi_3,\ell_4m_4\xi_4}
&\equiv&
{1 \over 4}~ \left\langle\ell_1m_1\xi_1,\ell_2m_2\xi_2\right| \hat V 
\left|\ell_3m_3\xi_3,\ell_4m_4\xi_4\right\rangle\nonumber\\ 
&\times& 
\sqrt{1+\delta_{\ell_1\ell_2}\delta_{m_1 m_2}\delta_{\xi_1\xi_2}}
\sqrt{1+\delta_{\ell_3\ell_4}\delta_{m_3 m_4}\delta_{\xi_3\xi_4}},
\end{eqnarray}
where $\hat V$ stands for the two-body terms
in the Hamiltonian $\hat H$.
It is worth noting that $\tilde \varepsilon$
is not diagonal in $\ell$. 

The energy~(\ref{Ener-ibm4}) depends
explicitly on the deformation parameters $\lambda$
and implicitly on $\alpha_T$, $\alpha_S$, and $\delta$
through $\tilde f_1$ and $\tilde f_2$.
The deformation parameters $\lambda$
are obtained by minimizing the energy
with the constraint of conserving the transformation norm,
\begin{equation}
\label{var-ener}
\delta[E(\lambda,\delta,\alpha_T,\alpha_S)-\sum_{\xi\ell m}E_\xi 
\lambda^{\xi *}_{\ell m}\lambda^{\xi}_{\ell m}]=0.
\end{equation}
The Hartree-Bose equations for IBM-4
are obtained by deriving
with respect to $\lambda^{\xi*}_{\ell m}$,
\begin{equation}
\label{Har2-ibm4}
\sum_{\ell_2m_2}
h_{\ell_1m_1,\ell_2m_2}^\xi
\lambda_{\ell_2m_2}^{\xi}=
E_\xi\lambda_{\ell_1m_1}^{\xi},
\end{equation}
where the Hartree-Bose matrix, $h^\xi$, is
\begin{eqnarray}
\label{Har3-ibm4}
\lefteqn{h_{\ell_1m_1,\ell_2m_2}^\xi
=\tilde\varepsilon_{\ell_1 m_1\xi \ell_2 m_2 \xi}
\tilde{f_1}(\delta,\alpha_T,\alpha_S,\xi\xi)
\delta_{m_1m_2}}\nonumber\\
&+&2\sum\limits_{\ell_3m_3\ell_4m_4\xi_2\xi_3\xi_4}
{V_{\ell_1m_1\xi,\ell_3m_3\xi_3,\ell_4m_4\xi_4,\ell_2m_2\xi_2}}
{\lambda_{\ell_3m_3}^{\xi_3*}\lambda _{\ell_4m_4}^{\xi_4}
 \lambda_{\ell_2m_2}^{\xi_2}
\over
{\lambda_{\ell_2m_2}^{\xi}}}
\tilde{f_2}(\delta,\alpha_T,\alpha_S,\xi\xi_3\xi_4\xi_2).\nonumber\\
\end{eqnarray}
There are six coupled equations of this form,
which are solved for fixed values of
$\alpha_T$, $\alpha_S$, and $\delta$
in a self-consistent way.
This procedure yields
the equilibrium deformation parameters $\lambda$.
The equilibrium values for the parameters  
$\alpha_T$, $\alpha_S$, and $\delta$
are obtained by an additional minimization.
In fact,
if the deformation parameters are independent of $\xi$,
the parameters $\alpha_T$ and $\alpha_S$
can be fixed to the value $-1/2$
which corresponds, as mentioned above,
to a state with well-defined spin and isospin.
As was shown for IBM-3~\cite{Gino94,Garc98}
this is a good approximation for $N=Z$ nuclei.

To test the present formalism,
comparisons with exact calculations are carried out.
Numerical calculations in the framework of IBM-4
are now possible~\cite{Juil99} but still difficult.
Also, only few dynamical limits have been studied.
Here the following schematic Hamiltonian is considered
\begin{equation}
\label{sch-ham-4}
\hat H= ~a~\hat C_2[SU_{TS}(4)]~+~b~\hat C_2[SU_{S}(3)]+
~c~\hat C_2[SU_{L}(3)],
\end{equation}
where $\hat C_2[G]$ stands
for the quadratic Casimir operator of the algebra $G$.
The first operator is an invariant of the algebra $SU_{TS}(4)$
which is the boson equivalent
of Wigner's supermultiplet algebra~\cite{Elli81}.
It is worth noting here that, as mentioned in \cite{Isac98}, there are
two alternative $SU_{TS}(4)$ limits with the same eigenspectrum but
different phases in the wave function. The results presented below  
are obtained within one them, which is associated to the election
of the operator $\hat{Y}_{\mu\nu}^+$ (using the notation of
Ref.~\cite{Isac98}). The use of the alternative limit, using
$\hat{Y}_{\mu\nu}^-$, gives identical results but changing sign to
$\delta$. 
The second operator in~(\ref{sch-ham-4})
is an invariant of the $SU_S(3)$ algebra
associated with the $\sigma$ ($S=1,T=0$) boson.
(Its definition is analogous to that of $SU_S(3)$
considered in the $L=0$ IBM-4 of Ref.~\cite{Isac98}.)
The last invariant in~(\ref{sch-ham-4})
is an orbital deformation term
associated with an $SU(3)$ algebra
which is scalar in spin and isospin.

With this Hamiltonian three situations are studied.
The first case corresponds to $a\neq0$ and $c=0$. 
In Ref.~\cite{Isac97} the competition of $T=0$ and $T=1$ pairing
was discussed using this Hamiltonian. This is a relevant test for the
formalism presented here since it explores the spin-isospin degrees of
freedom which makes the main difference of IBM-4 with respect to 
previous versions of IBM.
In this case the mean-field and exact calculations are almost identical
although the exact calculation is always slightly lower in energy. 
This can be 
appreciated in table~\ref{tab-1} where exact and mean-field
ground-state energies (in units of $a$) are given
for $N=5$ and $N=15$ bosons
for selected values of $b/a$.
The value $b=0$ yields
a Hamiltonian with the $SU(4)$ symmetry,
and degenerate lowest $T=0$ and $T=1$ states.
Negative values of the ratio $b/a$ favor $T=0$
while positive values favor $T=1$ pairing.
The expectation value
of the schematic Hamiltonian~(\ref{sch-ham-4}) with $c=0$
is independent of the deformation parameters. The minimum of the
energy occurs for $\xi$ independent parameters and 
$\alpha_T=\alpha_S=-{1\over 2}$.
This is so because it has no orbital dependence.
It is worth
noting that the variational wave functions
(\ref{Gswf-ibm4-pp}-\ref{Gswf-ibm4-ii2}) contain for special values of
$\delta$ the lowest eigenfunctions
of the $SU_T(3)\otimes SU_S(3)$ limit~\cite{Isac98}
($\delta=\infty$) and of the $SU_{TS}(4)$
limit~\cite{Isac98} ($\delta=- 1$). 

The second case corresponds to $a\neq0$ and $b=0$
and includes the deformation term $\hat{C}_2[SU_L(3)]$.
The exact ground-state energy is known analytically:
\begin{equation}
\label{Ener-su4-su3l}
E=a~ \lambda (\lambda +4) +~c~ 2N(2N+3). 
\end{equation}
This is so because,
for sufficiently large negative $c$,
the ground state belongs
to the $SU_{TS}(4)$ representation $(0,\lambda,0)$
and the $SU_L(3)$ representation $(2N,0)$,
where $N$ is the boson number
and $\lambda=T$ for even-even nuclei and $\lambda=1$ for
odd-odd $N=Z$ nuclei~\cite{Isac98}. The corresponding calculation 
is also performed with the mean-field formalism presented here and 
the exact results are reproduced. 
The calculation gives an intrinsic state with 
$\beta_\xi\equiv \lambda^\xi_{20}/\lambda^\xi_{00}=\sqrt{2}$,
$\alpha_T=\alpha_S=-{1\over 2}$ and $\delta=- 1$
which is an eigenstate of the Hamiltonian~(\ref{sch-ham-4}) 
with $a\neq0$ and $b=0$. 
This result is similar to that obtained for
the intrinsic state of IBM-1~\cite{Levi96}.

The last case considered is the general one with $a,~b,~c$ different
from zero. The ground state still belongs to the
$SU_L(3)$ representation $(2N,0)$ and hence the
contribution to the ground-state energy coming from $~c~ \hat
C_2[SU_{L}(3)]$ is diagonal. The other two terms in the
Hamiltonian can be diagonalized as in Ref.~\cite{Isac97}. Thus the
exact energies are those calculated in table~\ref{tab-1} 
(under `exact') plus
$c~ 2N(2N+3)$. This calculation is repeated with the mean-field
formalism and produces an intrinsic state with the same $SU_L(3)$
symmetry as the exact one, $(2N,0)$. As in the preceding case
$\beta_\xi\equiv\lambda^\xi_{20}/\lambda^\xi_{00}=\sqrt{2}$,
$\alpha_T=\alpha_S=-{1\over 2}$, but now $\delta\neq- 1$. The 
mean-field energies are those given in table~\ref{tab-1} 
(under `mean field') plus $c ~2N(2N+3)$.

These results demonstrate that 
the present mean-field formalism
has the correct ingredients 
for reproducing the full IBM-4 calculation.
In addition, this formalism allows calculations
for an arbitrary number of bosons
and a general Hamiltonian,
not necessarily corresponding to
a dynamical symmetry limit of the model.

To summarize, a Hartree-Bose mean-field approximation for IBM-4
has been presented,
along with trial wave functions
valid for even-even and odd-odd nuclei with $N=Z$.
The trial wave functions
include boson correlations in the spin and isospin spaces.
Comparisons with  exact calculations
show good agreement
from which can be inferred
that the present formalism gives
a good approximation to the full diagonalization.
The aim is now to consider more realistic IBM-4 Hamiltonians
that include both types of pairing
($T=0$ and $T=1$),
and a spin-orbit coupling
as well as more general quadrupole deformation terms.
This will enable the study of the interplay
between single-particle, spin-isospin,
and orbital degrees of freedom.
This work is currently in progress.

We wish to thank S.~Pittel and D.D.~Warner for useful comments. This
work has been supported by the Spanish DGICYT under contract
Nos. PB98-1111 and PB95-0123 and a DGICYT-IN2P3 agreement.



\begin{table}
\caption{Exact and mean-field energies
  of ground states, and their isospins, for selected values of $b/a$. 
  The cases shown correspond to $N=Z$ odd-odd nuclei with 
  $N=5$ and $N=15$ bosons, respectively.}
\begin{tabular}{rrrrcrr}
 &&\multicolumn{2}{c}{$E_{\rm gs}/a$ ($N=5$)}&&
   \multicolumn{2}{c}{$E_{\rm gs}/a$ ($N=15$)}\\
\cline{3-4}\cline{6-7}
$b/a$&$T$&exact&mean field&&exact&mean field\\
\hline
$-1.0$& 0   &$-27.8359$ &$-27.8357$ &&$-236.126$ &$-236.125$ \\
$-0.8$& 0   &$-20.5261$ &$-20.5257$ &&$-183.938$ &$-183.937$ \\
$-0.6$& 0   &$-13.4392$ &$-13.4383$ &&$-132.395$ &$-132.394$ \\
$-0.4$& 0   &$-6.69081$ &$-6.68917$ &&$-81.9874$ &$-81.9861$ \\
$-0.2$& 0   &$-0.46290$ &$-0.46140$ &&$-34.0488$ &$-34.0484$ \\
$ 0.0$& 0,1 &$ 5.00000$ &$ 5.00000$ &&$ 5.00000$ &$ 5.00000$ \\
$ 0.2$& 1   &$ 6.56989$ &$ 6.57374$ &&$ 11.8499$ &$ 11.9770$ \\
$ 0.4$& 1   &$ 7.76803$ &$ 7.78117$ &&$ 15.2339$ &$ 15.4644$ \\
$ 0.6$& 1   &$ 8.70462$ &$ 8.72895$ &&$ 17.5636$ &$ 17.8459$ \\
$ 0.8$& 1   &$ 9.45690$ &$ 9.49143$ &&$ 19.3676$ &$ 19.6743$ \\
$ 1.0$& 1   &$ 10.0763$ &$ 10.1187$ &&$ 20.8479$ &$ 21.1652$ \\
\end{tabular}
\label{tab-1}
\end{table}

\end{document}